\documentclass[12pt,preprint]{aastex}
\usepackage{epstopdf}
\usepackage{natbib}
\newcommand{\msun}{M$_{\odot}$}

\begin{document}

\title{A Substellar Companion to the Dusty Pleiades Star HD 23514} 

\author{David R. Rodriguez\altaffilmark{1,5},
Christian Marois\altaffilmark{2}, B.\ Zuckerman\altaffilmark{1}, Bruce Macintosh\altaffilmark{3}, Carl Melis\altaffilmark{4}}

\altaffiltext{1}{Dept.\ of Physics \& Astronomy, University of California, Los Angeles 90095, USA}
\altaffiltext{2}{National Research Council Canada, Herzberg Institute of Astrophysics, 5071 West Saanich Road, Victoria, BC V9E 2E7, Canada}
\altaffiltext{3}{Lawrence Livermore National Laboratory, 7000 East Avenue, Livermore, CA 94550, USA}
\altaffiltext{4}{Center for Astrophysics and Space Sciences, University of California, San Diego, CA 92093, USA}
\altaffiltext{5}{Current address: Departamento de Astronom{\'i}a, Universidad de Chile, Casilla 36-D, Santiago, Chile (drodrigu@das.uchile.cl)}

\begin{abstract}
With adaptive optics imaging at Keck observatory, 
we have discovered a substellar companion to the F6 Pleiades star HD~23514, one of the dustiest main-sequence stars known to date ($L_{IR}/L_{*}\sim2\%$).
This is one of the first brown dwarfs discovered as a companion to a star in the Pleiades.
The 0.06~M$_\odot$ late-M secondary has a projected separation of $\sim$360~AU.
The scarcity of substellar companions to stellar primaries in the Pleiades combined with the extremely dusty environment make this a unique system to study.
\end{abstract} 


\section{Introduction}

HD~23514 (HII~1132) is an F6 star in the Pleiades.
It is one of only a handful of youthful stars, ages 35 to $\sim$100 Myr, that are surrounded by warm dust particles located in the terrestrial planet zone and which absorb a percent or so of the bolometric luminosity of their central star \citep{Melis:2010}.
We have observed HD~23514 with an adaptive optics system at the Keck observatory using the NIRC2 camera to check the binarity of this unusually dusty star.

To the best of our knowledge, to date no substellar companion to a stellar primary has been found in the Pleiades. 
However, while this paper was being reviewed, \citet{Geissler:2011} announced the discovery of a substellar companion to the Pleiad HII~1348.
Studies of substellar objects in the Pleiades focus on 
free floating brown dwarfs or companions to low mass 
stars \citep{Stauffer:2007, Lodieu:2007, Bouy:2006, Martin:2003, Pinfield:2000}.
With an adaptive optics survey of 144 solar type stars in the Pleiades, \citet{Bouvier:1997} was 
sensitive  to objects with masses down to $\sim$0.03~M$_\odot$ at separations of $\sim$1\arcsec.
However, their lowest mass companion ($\sim$0.15~M$_\odot$)
is still above the substellar boundary (they did not observe HD~23514).
Our work has revealed one such substellar companion in the Pleiades.

\section{Observations}\label{observations}

We observed HD~23514 with the adaptive optics (AO) system and the NIRC2 narrow camera at the 10-m Keck telescope at Mauna Kea observatory. 
Observations were performed in natural guide star mode.
Some images were obtained in angular differential imaging mode (ADI) while others had the primary behind a 600~mas or 800~mas diameter coronographic spot.
The rest of the images were taken with dithered observations in position angle (PA) mode where the field of view is rotated so as to keep the sky fixed on the NIRC2 camera.
All data were reduced in the standard manner (flat fielding, dark subtraction, sky subtraction) with IRAF\footnote{IRAF is distributed by the National Optical Astronomy Observatory, which is operated by the Association of Universities for Research in Astronomy (AURA) under cooperative agreement with the National Science Foundation.} routines.
The data was linearized prior to processing using an IDL program\footnote{Written by S.\ Metchev, see http://www.astro.sunysb.edu/metchev/ao.html}.
We have applied the NIRC2 distortion solution presented in \citet{Yelda:2010} to refine our astrometry.
We adopt a pixel scale of $9.963 \pm 0.005$ mas/pix and a $0.13 \pm 0.02^{\circ}$ offset from North for the camera columns \citep{Ghez:2008}.
We did not apply any advanced LOCI processing to analyze the data.

Our observations are summarized in Table~\ref{tab:data}. 
Relative photometry was performed with IRAF's {\bf apphot} tasks using aperture sizes that maximized the signal-to-noise of the companion.
Generally, this was 1--1.4 times the FWHM of a point source.
The apparent magnitude for the companion at K$'$ is estimated by correcting the 2MASS Ks magnitude of $8.153\pm0.023$ for the primary as described by \citet{Wainscoat:1992}, namely K$'$--K $= 0.22$(H--K) and using the relationship between K and Ks described by \citet{Carpenter:2001}.
HD~23514 then has a K$'$ magnitude of $8.13\pm0.02$, not too different from the Ks magnitude.
For the coronographic data, the primary is attenuated by a factor of 28.57 and 27.80 for the 600 and 800~mas diameter spot, respectively (R.\ Galicher, private communication, 2011).
Multiple short exposures were obtained with integration times range from 10-60 seconds, as listed in Table~\ref{tab:data}, and the companion is clearly visible in most of the individual frames.
The primary is saturated in the 2006, 2007, and 2008 Ks and K$'$ long exposures. The individual frames were registered to shorter, unsaturated frames using IRAF's {\bf xregister} task. Photometry for these paired observations is performed with the same aperture size, but takes into account the different exposure times.

\begin{figure}[htb]
\begin{center}
\includegraphics[width=12cm,angle=0]{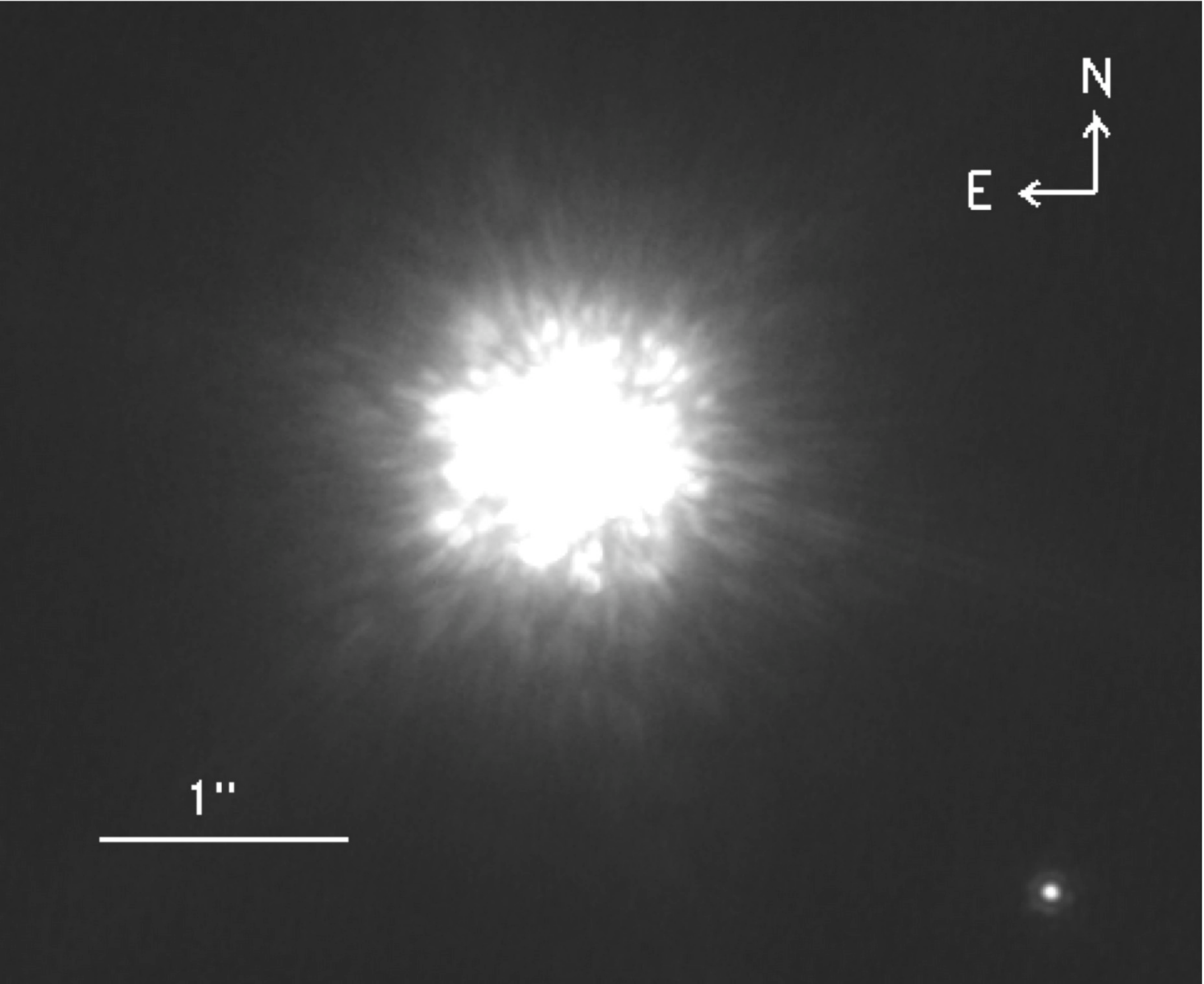}
\end{center}
\caption{HD~23514 observed with NIRC2 on November 4, 2008 (Ks filter).
The companion is located 2.6\arcsec~to the southwest of the primary star.
}
\label{fig:image}
\end{figure}

\begin{figure}[htb]
\begin{center}
\includegraphics[width=14cm,angle=0]{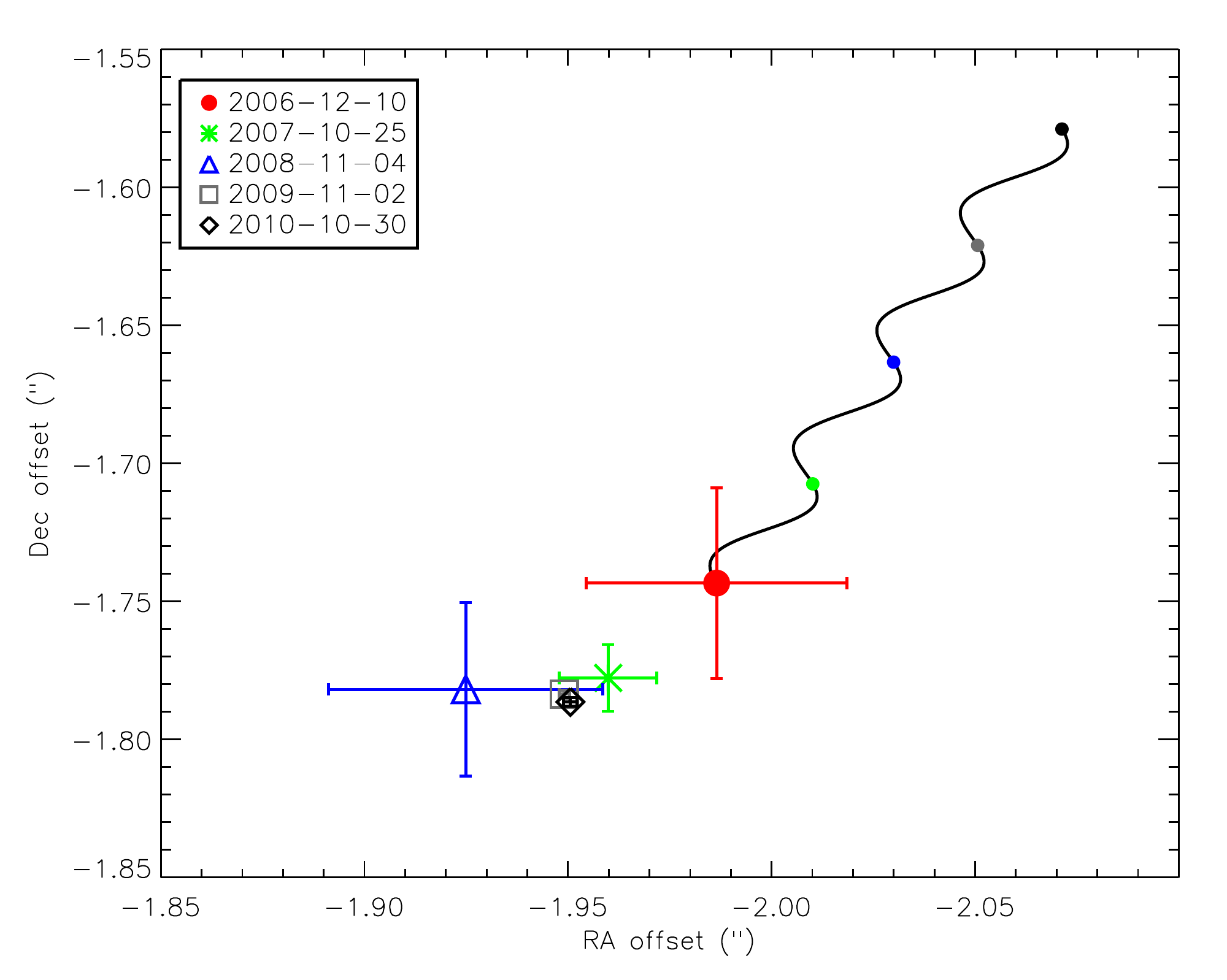}
\end{center}
\caption{Location of HD~23514B relative to the primary HD~23514. For 2009 and 2010 we use the measurements from the coronographic mode; see Table~\ref{tab:data}. The uncertainty in these two measurements are comparable to the symbol size and they effectively lie on top of each other in the figure. 
The black line indicates the relative motion a stationary background source would have throughout these observations with circles denoting the expected location at the time of our observations.
We adopt a distance of $135\pm10$~pc (see Section~\ref{results}).
}
\label{fig:location}
\end{figure}

We measure the candidate companion at a separation of $2.645\pm0.002$\arcsec~with position angle $227.52\pm0.02$\arcdeg~in our 2010 coronographic observations. The object has remained at the same location between epoch 2006 and 2010 (see Table~\ref{tab:meas} and Figure~\ref{fig:image}).
Had this been a more distant background object, then, based on the primary's proper and parallactic motion 
\citep[$20.5\pm0.4$ mas/yr, $-42.6\pm0.5$ mas/yr;][]{Zacharias:2010},
after 5 years this object should have been separated by $2.60\pm0.02$\arcsec~and 
had a PA of $232.7\pm1.0$\arcdeg~(see Figure~\ref{fig:location}).
As mentioned in Section~\ref{results}, the 2006 ADI data is somewhat unreliable as evidenced by the large uncertainties in Table~\ref{tab:meas}. Comparison from 2007 yields an expected position of a background source of $2.61\pm0.01$\arcsec~with PA of $230.78\pm0.29$\arcdeg.
Over this shorter 4-year baseline, a background object can be ruled out at the $\sim$12-$\sigma$ level.
The consistent values for separation and PA for the object from 2006 to 2010 imply this is a companion to the system or is a nearby co-moving Pleiades member, a possibility we address in Section~\ref{pleiad}.  

\begin{figure}[htb]
\begin{center}
\includegraphics[width=14cm,angle=0]{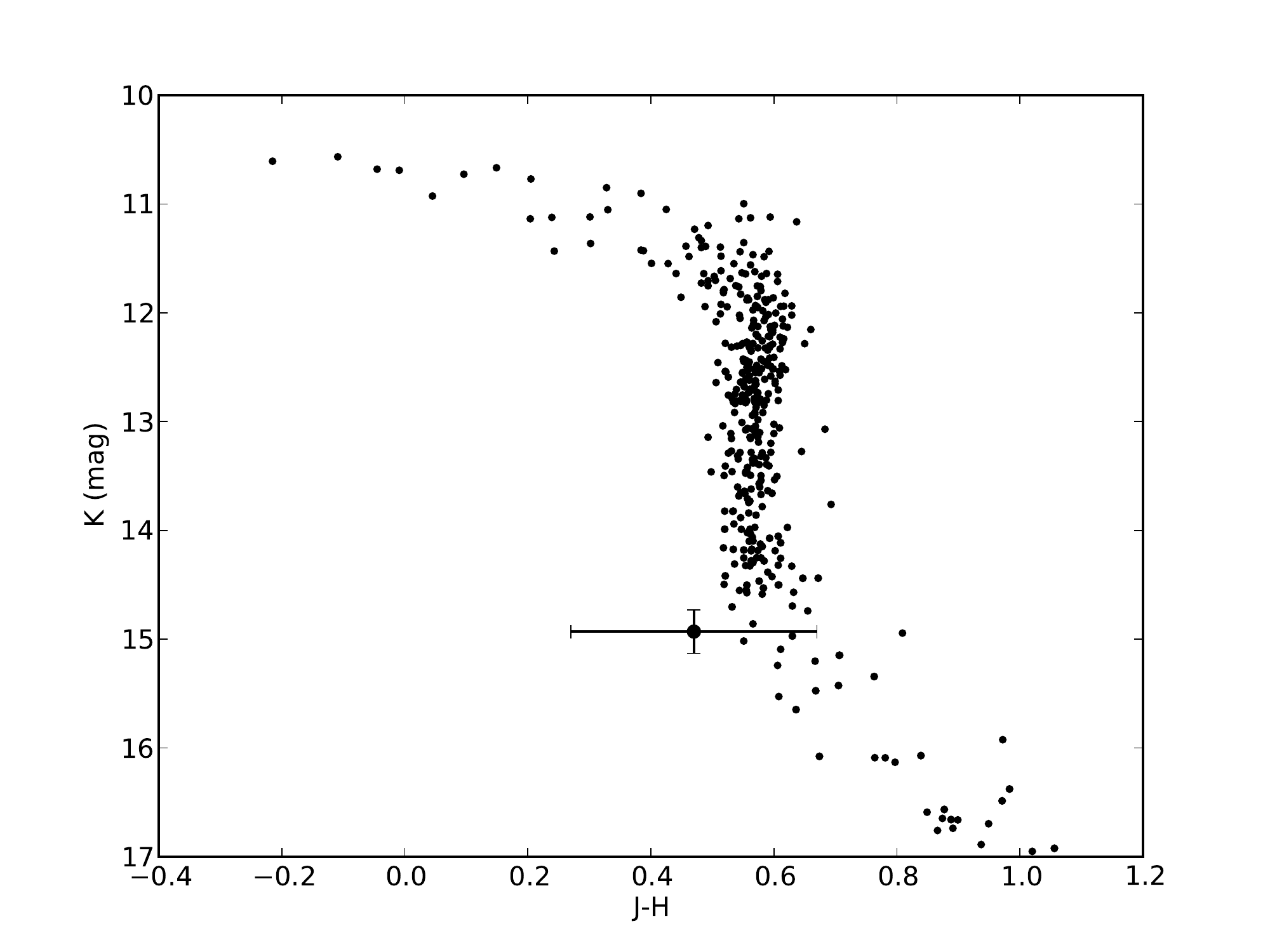}
\end{center}
\caption{Color magnitude diagram for Pleiades members listed in \citet{Lodieu:2007}. HD~23514 is the large circle with 0.2~magnitude error bars. Despite being bluer at J--H, HD~23514 is consistent with being a late-M object.
}
\label{fig:cmd}
\end{figure}

The quoted photometry and astrometry in Table~\ref{tab:meas} is the average of all the frames and the uncertainties are standard deviations of the individual measurements.
The more precise astrometry present in the latter epochs is possible since both the primary and secondary are detected and are unsaturated in the same frame.
A companion 6.8 magnitudes fainter than the primary F6 star at Ks would be a late-M dwarf (see Section~\ref{results}).
The J--Ks and H--Ks colors of this object are also consistent with being late-M ($\sim$M7--9; see \citealt{Kirkpatrick:1994}).
The J--H color is somewhat discrepant being bluer than expected; however, as shown in Figure~\ref{fig:cmd}, the measurement is consistent given the uncertainties.

\section{Results}\label{results}

Evolutionary models for a 100-Myr old F6 star like HD~23514 
\citep[1.35~M$_\odot$ and $T_{eff}\sim6400$~K;][]{Rhee:2008} predict 
an absolute Ks magnitude of about 2.5 \citep{Baraffe:1998,Baraffe:2002} 
giving a distance modulus of 5.65 and a distance of 135~pc.
An additional distance estimate can be obtained by considering that this star is part of the 
Pleiades, which lies 133.5~pc away (\citealt{An:2007,Soderblom:2005}; see also Table~3 in \citealt{vanLeeuwen:2009}).
The Pleiades extends nearly 10\arcdeg~in diameter 
\citep[and references therein]{vanLeeuwen:2009}; 
if we assume a spherical distribution, this amounts to a possible radial extent of 11.7~pc.
Hence, the distance to HD~23514 would be somewhere between 122~pc and 145~pc; we adopt a distance of $135\pm10$~pc. 
Considering our Ks photometry, the companion has $\Delta$Ks = $6.77\pm0.18$, which when comparing with the 2MASS photometry for HD~23514 gives an apparent magnitude of Ks = $14.92\pm0.18$. For our adopted distance, the absolute Ks magnitude of the companion is $9.27\pm0.24$.

To estimate the mass of the companion we make use of 
low-mass/substellar evolutionary models.
The DUSTY evolutionary models of \citet{Chabrier:2000} take 
into account the formation of condensates in the atmosphere and are appropriate 
for low-temperature objects down to $\sim$1300~K.
With these models, a 100~Myr object with absolute K magnitude 9.27 would have a mass somewhere 0.06 and 0.07 solar masses and effective temperature $\sim$2700~K. 
The evolutionary models of \citet{Baraffe:1998} use the NextGen dust-free atmospheres and are appropriate for objects with $T_{eff}\geq2300$~K.
These models similarly predict $T\sim2600$~K and mass $\sim$0.05--0.06 solar masses when compared to the Ks-band photometry.
The J and H-band photometry suggest somewhat lower masses (down to 0.05~solar masses) and cooler temperatures ($\sim$2500~K).
Combining all results together, we can infer that the companion has a mass of $0.06\pm0.01$ solar masses.
This is below the stellar/substellar boundary making the object a brown dwarf.
The temperature, likewise, appears to be $2600\pm100$~K, placing it at spectral type $\sim$M7 (see Figure~2 in \citealt{Reyle:2011}), consistent with the near-infared colors \citep{Kirkpatrick:1994}.

\begin{figure}[htb]
\begin{center}
\includegraphics[width=14cm,angle=0]{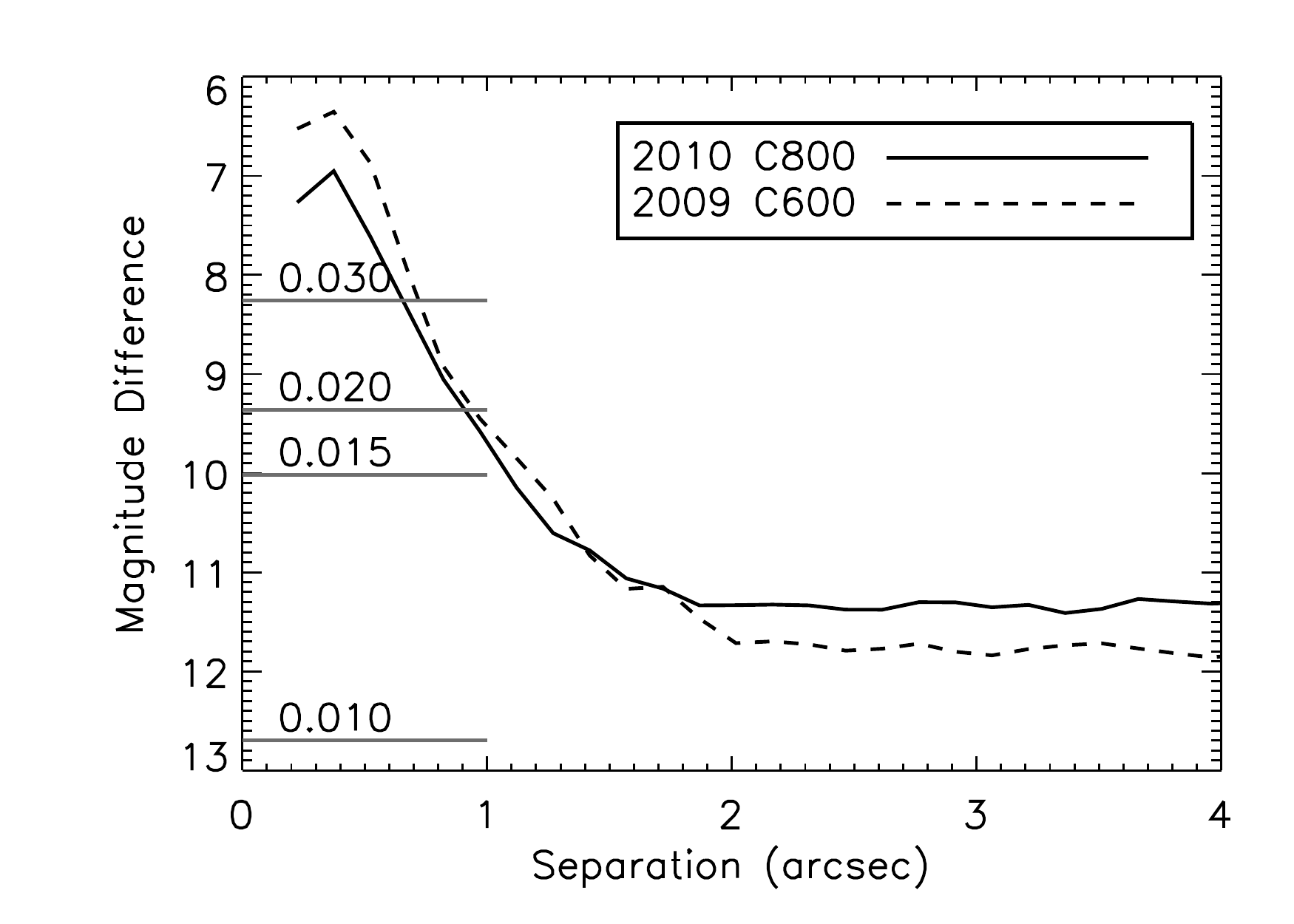}
\end{center}
\caption{Magnitude limits (5$\sigma$) as a function of separation for our 2010 and 2009 coronographic data. Limits are computed using the same aperture size by performing aperture photometry on blank regions of the sky in annuli around the primary. The limits are averaged over all frames. Companions $\sim$11-12 magnitudes fainter at Ks than the primary could have been detected at separations larger than 1.5\arcsec.
Mass limits in solar masses from the DUSTY evolutionary model of \citet{Chabrier:2000} are indicated.
}
\label{fig:limit}
\end{figure}

No other object is detected in the $10\arcsec \times 10\arcsec$ field of view. 
At the distance to HD~23514, this amounts to a radius of 675~AU.
Figure~\ref{fig:limit} presents the 5$\sigma$ magnitude limit as a function of distance from the primary for the two coronographic Ks observations (2009 and 2010). These are our best limits for HD~23514.
Limits are obtained by performing aperture photometry over blank regions of the sky at annuli surrounding the primary. These limits are averaged over the multiple frames.
No companion is detected with a limit of $\sim$11--12 magnitudes beyond 1.5\arcsec\, ($\sim$200~AU).
This suggests an absolute magnitude limit of $\sim$14 at Ks, or a mass of nearly $\sim$11 Jupiter masses when compared to evolutionary models \citep{Chabrier:2000,Baraffe:2003}.

The average Ks magnitude for the secondary, 14.92, differs from the K$'$ magnitude, 15.40, by half a magnitude.
A late-M star is expected to have H--Ks$\sim$0.5, which suggests K$'$--Ks$\sim$0.1, not enough to account the difference. 
However, the uncertainty in our K$'$ photometry is the highest of our measurements, at nearly 0.4 magnitudes. 
For the 2006 data, the uncertainty is a result of the larger aperture size ($\sim2.4\times$ FWHM) needed as the M-type companion is smeared out in the ADI exposures. 
For 2007 and 2008, the primary is saturated and the secondary's flux (and location) is compared to separate, shorter exposures as described in Section~\ref{observations}.
The high uncertainty, combined with the expected difference in the K$'$ and Ks filters is enough to account for the discrepancy.

\subsection{Is HD~23514B a free-floating Pleiad?}\label{pleiad}

While the common motion with HD~23514A is suggestive, there exists the possibility that the observed object may be a more distant (or closer) free-floating Pleiades brown dwarf.
As a member of the Pleiades, it would have about the same proper motion and thus appear to be co-moving with HD~23514.
The most robust way to ascertain this would be to observe the system over many years in order to detect (or rule out) orbital motion.
However, statistical arguments suggest the brown dwarf is indeed bound to HD~23514.

\citet{Jameson:2002} used a group of surveys to estimate the spatial distribution of brown dwarfs in the Pleiades.
HD~23514 is located about 1.7 degrees from the center ($03^h 47^m$, $+24^\circ 07'$; \citealt{Pinfield:1998}).
At that distance, \citet{Jameson:2002} suggest a brown dwarf (mass $\geq0.05$~\msun) density of 2.7 per square degree.
However, more recent studies performed by \citet{Lodieu:2007} and \citet{Moraux:2003} suggest average brown dwarf (mass $\geq0.03$~\msun) densities of 7.8 and 6.3 per square degree, respectively. \citet{Moraux:2003} surveyed the Pleiades in fields located 0.75 to 3.5 degrees away from the field center, \citet{Lodieu:2007} incorporated that survey and in addition also covered the center of the Pleiades with UKIDSS.
Thus, the area covered by these deeper surveys is similar in radial extent as that in \citeauthor{Jameson:2002} (\citeyear{Jameson:2002}; out to 2.2~degrees from the center), but the density of brown dwarfs is larger by a factor of about 2 than the average brown dwarf density of 4.2 per square degree found by \citet{Jameson:2002}.
To account for these deeper surveys (mass $\geq0.03$~\msun), we scale the density provided by \citet{Jameson:2002} and find that, at HD~23514's distance from the center of the Pleiades, the density of brown dwarfs is about $\sim$5 per square degree.
The NIRC2 field of view is $10\arcsec\times10\arcsec$, or an area of $7.7\times10^{-6}$ square degrees.
The likelihood of finding a free-floating Pleiades brown dwarf with mass $\gtrsim0.03$\msun\, in the field of view is thus about $4\times10^{-5}$.

The small likelihood derived lends weight to the idea this is a bound companion.
There are few brown dwarf companions known to stellar primaries (the well-known ``brown dwarf desert" phenomenon; \citealt{McCarthy:2004,Metchev:2009}).
Even though the frequency of brown dwarf companions at separations $\gtrsim$30~AU is small -- estimated to range from 0.5\% to nearly 20\% (see discussion in \citealt{Kraus:2008}) -- HD 23514B is more likely to be a bound companion than an isolated Pleiades brown dwarf.

If we assume the orbit is circular and face-on, we can use the measured projected separation to estimate the yearly orbital motion of the system. Given a mass of 1.35\msun, we estimate a period of $\sim$6000~years. This yields an angular motion of 0.06 degrees per year or a linear motion of $\sim$2.8~mas per year. This is comparable to our astrometric precision in our later epochs (see Table~\ref{tab:meas}), suggesting a baseline of at least 5 more years would be needed to confirm orbital motion. However, the yearly motion is likely to be smaller as the orbit may be inclined and/or eccentric.

\begin{figure}[htb]
\begin{center}
\includegraphics[width=14cm,angle=0]{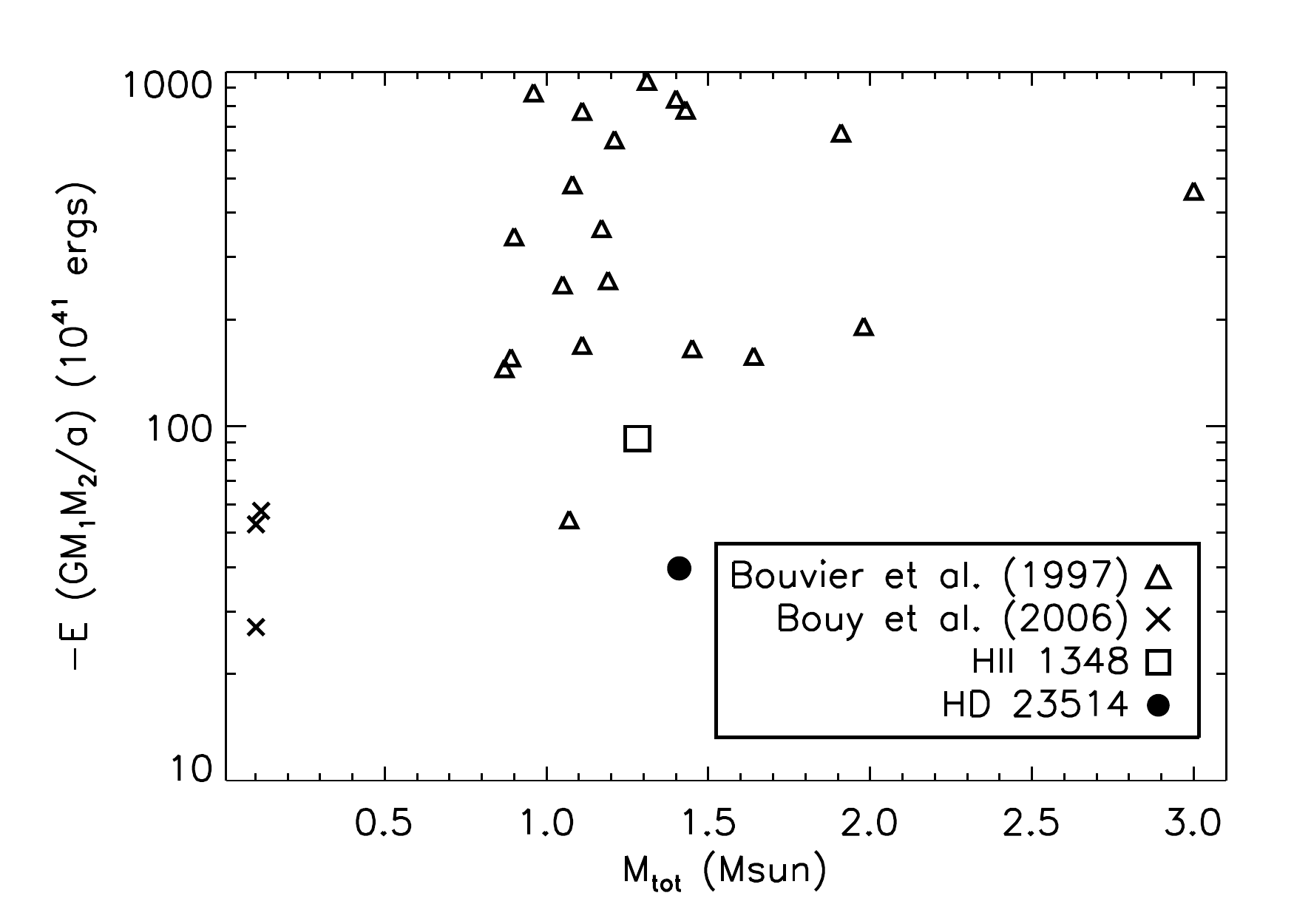}
\end{center}
\caption{Binding energy of resolved binaries in the Pleiades. While HD~23514 does have a low binding energy, there are substellar binaries with comparable or even lower values.
HII~571 (0.89+0.18\msun; 518~AU) is the other stellar binary with low binding energy.
\citet{Geissler:2011} recently announced the discovery of a substellar companion to HII~1348.
}
\label{fig:binde}
\end{figure}

It is instructive to consider whether or not a substellar object at that separation will remain bound over the 100~Myr lifetime of the Pleiades.
Figure~\ref{fig:binde} presents the binding energy of resolved stellar and substellar binaries in the Pleiades (from \citealt{Bouvier:1997,Bouy:2006}).
Note that since inclination and eccentricity are typically not available for these systems (and certainly not for HD~23514), the true binding energy may be somewhat different than plotted here.
Despite not being the most widely separated binary in the Pleiades (\citeauthor{Bouvier:1997}\ present systems at $\sim$900~AU), HD~23514 does have the lowest binding energy among the stellar mass binaries. However, substellar Pleiades binaries with separations $\lesssim$20~AU have comparable or lower binding energies.
The similar binding energies suggests that stellar interactions within the Pleiades, while more frequent than in the field, may not be enough to disrupt the HD~23514 system in timescales much shorter than 100~Myr.
We note that HIP~78530 \citep{Lafreniere:2011} and 1RXS~J160929.1--210524 \citep{Lafreniere:2010,Lafreniere:2008} both host substellar companions at wide separations in the 5~Myr-old Upper Scorpius region. These are considered physically bound based on common proper motion, similar to the case of HD~23514. 
However, both systems have estimated binding energies lower than that of HD~23514. 

\section{Discussion}\label{discussion}

Previous searches for brown dwarf companions in the Pleaides have focused on 
low-mass stars or free-floating brown dwarfs \citep[e.g.,][]{Stauffer:2007,Lodieu:2007,Bouy:2006,Martin:2003,Pinfield:2000}.
\citet{Bouvier:1997} used adaptive optics to observe 144 G- and K-type stars in the Pleiades.
The lowest mass companion they found was 0.15~M$_\odot$, though they were sensitive to objects with masses down to $\sim$0.03~M$_\odot$ at separations of $\sim$1\arcsec.
A deeper study conducted by \citet{Metchev:2009} also looked at solar-type stars, but found no substellar companions among $\sim$20 Pleaides stars.
Hence, while there may be hundreds of free-floating brown dwarfs in the Pleaides \citep{Lodieu:2007,Jameson:2002}, this is the first brown dwarf detected as a companion to a sun-like star.
The scarcity of companion brown dwarfs to the $\sim$300 solar-type stars in the Pleiades \citep{Stauffer:2007,Pinfield:1998} is consistent with the findings of a brown dwarf desert among solar-type stars out to separations of $\sim$1000~AU \citep{McCarthy:2004,Metchev:2009}.

HD~23514 is surrounded by a substantial amount of warm ($\sim$700~K) dust in a thin ring located $\sim$0.25~AU from the star if the emission comes from large blackbody grains or at most a few AU if the grains are smaller and radiate less efficiently \citep{Rhee:2008}.
At a distance of 135~pc, the companion in Figure~\ref{fig:image} has a projected separation of about 360~AU, far larger than the dust semi-major axis.
Only a handful of systems are known to contain substantial quantities ($L_{IR}/L_*\gtrsim0.1\%$) of warm dust: Cha-Near member EF~Cha \citep{Rhee:2007b}, $\beta$~Pic member HD~172555 \citep{Lisse:2009}, Lower Centaurus Crux member HD~113766 \citep{Chen:2006}, AB Dor member HD~15407 \citep{Melis:2010}, M47 member P1121 \citep{Gorlova:2004}, and NGC2547 member ID8 \citep{Gorlova:2007}.
BD+20~307 is another main-sequence star known to contain a large amount of warm dust \citep{Song:2005}, however it is much older ($\sim$1~Gyr) than these other systems \citep{Zuckerman:2008b,Weinberger:2008}.
Of these 8 systems, HD~23514, HD~15407, HD~172555, BD+20~307, and HD~113766 are known to be binary or triple systems.
All are wide separation systems ($>$100~AU) with the exception of BD+20~307, whose separation is only about 0.05~AU (though with a high probability of a third more distant, and as yet unseen, object; see \citealt{Tokovinin:2006}).
Interestingly, HD~23514, HD~15407, and HD~172555 all exhibit signs of silica emission suggesting high-velocity impacts produced the dust in the system \citep{Melis:2010, Lisse:2009}.

Binaries with separations of a few tens of AU are less likely to contain substantial amounts of dust when compared with single stars or with wide separation binaries \citep{Trilling:2007,Rodriguez:2011}.
However, if HD~23514 hosts a tertiary companion, say a close-in planet, whose initial inclination is different than that of the substellar companion, the Kozai mechanism \citep{Kozai:1962} can work to disrupt its orbit by continuously changing the inclination and eccentricity. This can potentially stir up planetesimals in the system and thus enhance collisions in the terrestrial zone (for example, see \citealt{Malmberg:2007,Innanen:1997}).
\citet{Rhee:2008} suggests a catastrophic or even planet-planet collision to account for the large amount of dust in the system. Such an event could take place if the eccentricity of a planet or large planetesimal is sufficiently increased through the Kozai mechanism. A large eccentricity can lead to high-velocity impacts near periastron, which can account for the silica emission observed in the system.
Equation~36 in \citet{Ford:2000} can be used to estimate the period of the eccentricity oscillation of this tertiary body. For a planet or planetesimal at 5~AU, and assuming HD~23514B's orbit is circular and its projected separation is the semi-major axis, the period is $\sim$80~Myr. While there are many unknowns, it is interesting to note that this is comparable to the age of the system. Further monitoring of the system, as well as deeper searches for any additional companions closer to the star, will be key in determining the relationship, if any, between the secondary companion and the dust around the primary.

\section{Conclusions}\label{conclusions}

Using adaptive optics imaging at Keck, we have discovered a substellar companion to the dusty Pleiades star HD~23514.
Based on comparison with evolutionary models, we estimate the companion to have a mass of about $0.06\pm0.01$\msun\, and temperature $2600\pm100$~K.
To our knowledge, HII~1348 and HD~23514 are the only Pleiades stars known to have brown dwarf companions \citep{Geissler:2011}.

HD~23514 is one of the dustiest main-sequence stars known to date with $L_{IR}/L_{*}\sim2\%$.
The warm dust present in the system is located within a few AU of the primary.
In contrast, the companion has a projected separation of $\sim$360~AU.
At that distance, the companion's gravity is expected to have little influence over the evolution of the warm dust seen around the primary. However, if a tertiary object were present in the system very close to the star, the Kozai mechanism can affect its orbit and possibly enhance collisions that generate dust.

\acknowledgements
{\it Acknowledgements.} We thank J.\ Stauffer for useful discussions on the Pleiades, R.\ Galicher for the NIRC2 coronographic attenuation, and S.\ Yelda for assistance in NIRC2 reduction. 
We thank our anonymous referee for a prompt review and constructive suggestions.
The data presented herein were obtained at the W.M. Keck Observatory, which is operated as a scientific partnership among the California Institute of Technology, the University of California and the National Aeronautics and Space Administration. The Observatory was made possible by the generous financial support of the W.M. Keck Foundation.
The authors wish to recognize and acknowledge the very significant cultural role and reverence that the summit of Mauna Kea has always had within the indigenous Hawaiian community.  We are most fortunate to have the opportunity to conduct observations from this mountain.
This research was supported in part by NASA grants to UCLA.



\begin{deluxetable}{llccccccc} 
\tabletypesize{\footnotesize}
\tablecolumns{7}
\tablewidth{0pc}
\tablecaption{HD23514 Observations \label{tab:data}}

\tablehead{
\colhead{UT Date} & \colhead{Filter} & \colhead{Integration} & \colhead{N$_{coadds}$} & \colhead{Exposure} & \colhead{N$_{frames}$} & \colhead{Notes} \\
\colhead{} & \colhead{} & \colhead{Time (s)} & \colhead{} & \colhead{Time (s)} & \colhead{} & \colhead{} }

\startdata
2006-12-10	& K$'$ & 30 & 2 & 60 & 18 & ADI mode, primary saturated \\
2006-12-10	& K$'$ & 0.2 & 100 & 20 & 9 & ADI mode, secondary not detected\tablenotemark{a} \\
2006-12-10	& K$'$ & 0.005 & 50 & 0.25 & 12 & ADI mode, secondary not detected\tablenotemark{b} \\
2007-10-25	& K$'$ & 30 & 2 & 60 & 14 & primary saturated\\
2007-10-25	& K$'$ & 0.018 & 20 & 0.36 & 8 & secondary not detected\tablenotemark{a} \\
2007-10-25	& K$'$ & 0.005 & 50 & 0.25 & 6 & secondary not detected\tablenotemark{b} \\
2008-11-04	& Ks & 20 & 1 & 20 & 8 & primary saturated\\
2008-11-04	& Ks & 0.5 & 20 & 10 & 15 & secondary detected in 9/15 frames\\
2009-11-01	& H & 1 & 60 & 60 &	7 & \\
2009-11-02	& Ks C600 & 30 & 1 & 30 & 14 & coronographic; 600~mas diameter spot \\
2010-10-30	& J & 0.053 & 200 & 10.6 & 5 & \\
2010-10-30	& H & 0.053 & 200 & 10.6 & 10 & \\
2010-10-30	& Ks	& 0.1 & 200 & 20 & 5 & \\
2010-10-30	& Ks C800 & 6 & 2 & 12 & 10 & coronographic; 800~mas diameter spot \\
\enddata

\tablenotetext{a}{The full chip ($1024\times1024$ pixels) was not read out, only the central $256\times264$ pixels. The secondary is not in the field of view.}
\tablenotetext{b}{The full chip was not read out, only the central $64\times120$ pixels. The secondary is not in the field of view.}

\tablecomments{
All data were taken with the NIRC2 narrow camera at the 10-m Keck telescope.
Data not labeled as ADI or coronographic were taken as dithered observations in PA mode.
The short exposure frames where the secondary is not detected are used to register those in which the primary is saturated.
Values quoted in Table~\ref{tab:meas} are averages over all frames.
}

\end{deluxetable}

\begin{deluxetable}{llccccccc} 
\tabletypesize{\footnotesize}
\tablecolumns{7}
\tablewidth{0pc}
\tablecaption{HD23514 Measurements \label{tab:meas}}

\tablehead{
\colhead{UT Date} & \colhead{Filter} & \colhead{Aperture} & \colhead{$\Delta$m} & \colhead{Apparent} 
& \colhead{Separation} & \colhead{PA} \\
\colhead{} & \colhead{} & \colhead{Size (mas)} & \colhead{(mag)} & \colhead{mag (mag)} 
& \colhead{(\arcsec)} & \colhead{(\arcdeg~E of N)} }

\startdata
2006-12-10	&	K$'$	&	45	& $	7.25	\pm	0.42	$ &	$15.38\pm0.42$	& $	2.64	\pm	0.02	$ & $	228.7	\pm	1.0	$ \\
2007-10-25	&	K$'$	&	25	& $	7.19	\pm	0.34	$ &	$15.32\pm0.34$	& $	2.64	\pm	0.01	$ & $	227.8	\pm	0.3	$ \\
2008-11-04	&	Ks	&	20	& $	6.66	\pm	0.33	$ &	$14.81\pm0.33$	& $	2.62	\pm	0.04	$ & $	227.2	\pm	0.5	$ \\
2009-11-01	&	H	&	25	& $	7.32	\pm	0.08	$ &	$15.61\pm0.08$	& $	2.642	\pm	0.003	$ & $	227.51	\pm	0.04	$ \\
2009-11-02	&	Ks C600	& 20	& $	6.59	\pm	0.22	$ &	$14.74\pm0.22$	& $	2.642	\pm	0.001	$ & $	227.54	\pm	0.03	$ \\
2010-10-30	&	J	&	25	& $	7.44	\pm	0.12	$ &	$15.92\pm0.12$	& $	2.644	\pm	0.004	$ & $	227.5	\pm	0.1	$ \\
2010-10-30	&	H	&	25	& $	7.10	\pm	0.05	$ &	$15.39\pm0.06$	& $	2.644	\pm	0.002	$ & $	227.48	\pm	0.05	$ \\
2010-10-30	&	Ks	&	25	& $	6.70	\pm	0.08	$ &	$14.85\pm0.08$	& $	2.642	\pm	0.005	$ & $	227.47	\pm	0.09	$ \\
2010-10-30	&	Ks C800	& 25	& $	6.90	\pm	0.09	$ &	$15.05\pm0.09$	& $	2.645	\pm	0.002	$ & $	227.52	\pm	0.02	$ \\
\enddata

\tablecomments{
All data were taken with the NIRC2 camera at the 10-m Keck telescope, as listed in Table~\ref{tab:data}.
The uncertainties are standard deviations of multiple frames; for the apparent magnitude, this error is added in quadrature with the primary's photometric error ($\sim$0.02 mag).
Aperture sizes were chosen to optimize the signal-to-noise of the secondary and correspond to roughly $1-1.4\times$ FWHM for a point source.
In the ADI data (2006), a larger aperture size was used ($\sim2.4\times$ FWHM) as the rotation of the field of view smeared out the companion (no advanced ADI/LOCI processing was used to analyze the data).
}

\end{deluxetable}

\clearpage

\end{document}